# All-charm tetraquarks

Richard J. Lloyd and James P. Vary

Iowa State University, Department of Physics and Astronomy

We investigate four-body states with only charm quarks. Working in a large but finite oscillator basis, we present a net binding analysis to determine if the resulting states are stable against breakup into a pair of $c\bar{c}$ mesons. We find several close-lying bound states in the two models we examine.

## Introduction

The spectrum of multiquark states has been investigated for three decades, but very little in the way of quantitative results for the spectrum of all-charm tetraquarks, $c^2\bar{c}^2$, has been published. Iwasaki[1] provides the first treatment that argues bound states of $c^2\bar{c}^2$ exist and estimates their mass. His argument is based on a string model of hadrons and a generalized OZI rule to establish the plausibility of $c^2\bar{c}^2$ exotics. The best value for the charm quark mass of 1.5 GeV (in 1975) was used to estimate the mass of the tetraquark bound state in the neighborhood of 6 GeV (or 6.2 GeV if the $J/\psi$ was used).

A number of other approaches to multiquark models exist in the literature. Jaffe[2], using an early version of the bag model, made predictions for tetraquark spectroscopy of the type $q^2\overline{q}^2$ with q a light quark (lighter than charm). In a similar vein, Schaffner-Beilich and Vischer[3] give impressive systematic treatments of charmlets (systems with at least one charm quark plus lighter quarks) but do not treat $c^2\overline{c}^2$.

In addition to the bag model, many variants of a wide class of potential models are used to describe multiquark states. Successful models founded upon the work of Isgur and Weinstein[4] tended to confirm the earlier prediction of Jaffe, and identified the $a_0(980)$and $f_0(980)$as molecular states of four quarks, for example (see especially Isgur and Godfrey[5] ). Models in the spirit of Karl and Ericson[6] employ a pion exchange potential to investigate hadronic molecules, including meson-meson, meson-baryon and baryon-baryon systems. Stancu[7] investigated the stability of multiquark hadrons and compared the one gluon exchange potential approach with the Goldstone boson exchange models of Glozman[8] and found, for example, that the models differed on the question of the existence of the H-dibaryon. However, all these models include at least one light quark in the bound state and do not consider four bound charmed quarks.

Lattice methods represent another approach to many body states. A massive and comprehensive review of the state of the art of lattice efforts by Bali[9] is worthwhile and instructive reading. To our knowledge, however, lattice methods have not been applied to heavy tetraquark bound states of the type we examine here.

Most recently, for example, the report by BaBar[10] of a narrow state at 2.32 GeV has stimulated a large number of theoretical efforts (see Barnes, Close and Lipkin[11] and references therein) with tetraquark states that include two flavors of light quarks. Tetraquarks of a single heavy flavor have not received similar attention although experimentally, double charmonium production has been reported at Belle[12] and BaBar[13] at center of mass energy over 10 GeV. One motivation for the present effort is to examine whether similar experiments designed to detect all-charm tetraquarks could be theoretically motivated.

In the remainder of this paper, we present a pair of parameterized Hamiltonians that are used to approximately fit the low-lying charmonium spectrum. These Hamiltonians are then employed to calculate the spectrum of all-charm tetraquarks in a quantum 4-body framework. We then present a net binding analysis to determine if bound states are obtained in our treatment.

## The Hamiltonian

The Hamiltonians we employ are non-relativistic with interactions inspired by the one gluon exchange potential. The main difference is that we treat the coupling strengths of the interactions as free parameters and simply adjust them to achieve an acceptable fit to the low-lying charmonium system. The specific form for our Hamiltonian is

$$H = T_{rel} + \frac{1}{2}\kappa\vec{\lambda}^2 + \alpha\left(H_{cm} - \frac{3}{2}\hbar\Omega\right) + \sum_{i<j}\vec{\lambda}_i \bullet \vec{\lambda}_j \left[\frac{a}{r_{ij}} - \sigma r_{ij} + \beta_0 \delta^3(r_{ij})_{S=0} + \beta_1 \delta^3(r_{ij})_{S=1} - \omega V_{tens} - \eta V_{so}\right] \tag{1}$$

The second and third terms of the Hamiltonian serve as projection operators. The first operator, a simple multiple of the SU(3) color Casimir operator acts on non-singlet color states so that, with positive $\kappa$ , it pushes them up relative to the color singlets. The second is a simple harmonic oscillator Hamiltonian that acts on the center of mass (CM) coordinates so that, with positive $\alpha$ , it pushes states of excited CM motion higher in the spectrum relative to the states with pure 0S harmonic oscillator CM motion. For all applications in this paper we will set $\kappa$ and $\alpha$ equal to 3 GeV and 10 (unitless), respectively. These values are sufficiently large to provide a clean separation between the low-lying states we present here, the “physical” states, and states with color non-singlet character and/or excited CM motion.

The tensor and spin-orbit potentials are defined by the following relationships. The orbital angular momentum referred to in the spin-orbit interaction is the relative orbital angular momentum.

$$V_{tens} = \frac{1}{r^3}\left[\vec{S}_1 \bullet \vec{S}_2 - 3\vec{S}_1 \bullet \hat{r}\vec{S}_2 \bullet \hat{r}\right] \qquad V_{so} = \frac{\vec{l} \bullet \vec{S}}{r^3} \tag{2}$$

The other parameter values in the Hamiltonian will be specified with each of the two fits.

We realize that our Hamiltonian admits van der Waals interactions in a multiquark system. Willey[14] has estimated the long-range contribution of these forces to behave as $R^{-7}$ where R is the separation of two color singlet subsystems. An earlier work by Greenberg and Lipkin[15] estimates a different power law dependence that is inconsistent with experimental data. Working in the limited spaces that we employ, we hope that such long-range forces do not affect our conclusions.

## The computational framework and results

To solve the quantum 2-body and 4-body problems, we select a single particle basis of harmonic oscillator states and construct Slater determinants of good total magnetic projection and good parity. We diagonalize the Hamiltonian in the Slater determinant basis to obtain the mass spectra and wavefunctions using the Lanczos algorithm[16]. From the wavefunctions, we evaluate the total angular momentum, J, of each eigenstate.

The basis depends on two parameters. One parameter, N, determines the dimension of the basis space. N defines the maximal number of allowed oscillator excitations in a given many body basis state by imposing the condition that $\sum_i (2n_i + l_i) \leq N$, where the sum is over the single particle oscillator quanta, $2n + l$, present in that state. The second parameter, $\hbar\Omega$, denotes the oscillator energy which we employ as one of our adjustable parameters fit to the charmonium spectrum.

This many-body computational framework has been used in other multi-fermion applications[17]. Those previous applications also featured a derivation of an effective Hamiltonian that we do not carry out here. Instead, we adapt the established procedures for many-body basis space enumeration, many-body Hamiltonian evaluation, and Lanczos diagonalization for the present effort. We pay special attention to the treatment of the color degree of freedom. That is, we impose the restriction that our physical many-fermion states are global color singlets with the projection method described above.

We first fix our Hamiltonian and basis space parameters by fitting the low-lying charmonium spectra. Two fits are obtained so we may begin to explore the sensitivity of our $c^2\bar{c}^2$ results to variations in acceptable $c\bar{c}$ fits. We select a maximum N, called $N_{\max}$=9 (10) for our negative (positive) orbital parity 2-body states in which we fit our parameters to the lowest few experimentally determined states. We then solve the 4-body problem in a sequence of model spaces by varying N up to $N_{\max}$. The maximum matrix dimension encountered in the 4-body problem was 3,013,782.

For each set of Hamiltonian parameters, designated Case 1 and Case 2 below, the sensitivity of the 2-body charmonium spectrum to spaces of smaller N values indicates that the parameters would require adjustment if $N_{\max}$ is changed.

**Method of tetraquark net binding analysis**

Since our basis space is comprised of harmonic oscillator states, our spectrum will exhibit an appearance of being bound, but this is not sufficient to conclude that a solution actually corresponds to a physically bound state. Qualitatively, we may view each 4-body state as composed of a pair of 2-body subsystems with a relative kinetic energy between them. Assuming the pairs will break apart, we can identify the minimum possible kinetic energy between the pairs based on the size of the model space, N. Let us call this minimum possible kinetic energy between the pairs as the "kinetic energy penalty", the penalty of confining two pairs of free particles in a finite harmonic oscillator basis. By "free particle" we mean the solution of our Hamiltonian with 2-body relative interaction terms (Eq. (1) summation term) removed. We quantify this penalty in Eq. (3).

$$\Delta M_{cc\bar{c}\bar{c}}^{free,N,\hbar\Omega} = M_{cc\bar{c}\bar{c}}^{free,N,\hbar\Omega} - \underset{n}{Min}\left[M_{c\bar{c}}^{free,n,\hbar\Omega} + M_{c\bar{c}}^{free,N-n,\hbar\Omega}\right] \qquad (3)$$

$M^{free,N,\hbar\Omega}$ denotes the lowest mass of the indicated (subscript) free particle system in the basis space specified by N and $\hbar\Omega$. Note that we subtract the minimum relative kinetic energy of a pair of color singlet 2-body subsystems from the minimum relative kinetic energy of the color singlet 4-body state. This correction tends to zero as $N_{\max}$ increases.

When we compute the spectrum of the full Hamiltonian in the 4-body system, we will then subtract the penalty in Eq. (3) from the results as follows.

$$M_{cc\bar{c}\bar{c}}^{corrected,N,\hbar\Omega} = M_{cc\bar{c}\bar{c}}^{full,N,\hbar\Omega} - \Delta M_{cc\bar{c}\bar{c}}^{free,N,\hbar\Omega} \qquad (4)$$

To be specific, we subtract the same kinetic energy penalty from every tetraquark mass eigenvalue of the "full" Hamiltonian to arrive at the "corrected" mass of each state regardless of its quantum numbers. The tetraquark spectrum, corrected via Eq. (4), will then be compared to charmonium pairs from the 2-body spectrum to determine if there is net binding of the 4-body state.

**Case 1 tetraquark results**

We now present the results of our first fit to the lowest 4 states for each parity in the charmonium spectrum, "Case 1", in the $N_{\max}$=9 and 10 model spaces. We also present our tetraquark results with the same Hamiltonian.

Tables 1 and 2 show the even and odd orbital parity states of charmonium for the full space of the fit, $N_{\max}$, as well as for spaces N< $N_{\max}$. The results in truncated spaces will be used in the tetraquark threshold analysis below. The experimental masses are taken from the Particle Data Group compilation[18]. The fit of the positive orbital parity states in Table 1 is only fair. This is due mainly to the fact that the delta functions are crude approximations to the short distance interactions. This aspect of our Hamiltonians limited our ability to simultaneously fit both the 1S and 2S splittings of the positive orbital parity states. We decided to adjust the delta functions to fit the $J/\psi$ and the $\eta_c(2S)$ masses in order to represent states with a range of radial excitations at the 4-body level. The root mean square (RMS) mass difference between the N=10 results and the experimental masses is 106 MeV for all four states, but falls to 54 MeV if one ignores the $\psi(2S)$. The RMS mass difference for the N=9 negative orbital parity results in Table 2 is 10.4 MeV, an excellent fit. If one uses all 8 states from both parities, the overall RMS mass difference is 75.3 MeV. For comparison, a recent relativistic constituent quark model based upon a covariant constraint dynamics approach[19] provides an RMS mass difference of 19.8 MeV for these same 8 states. For consistency, we will use our calculated $c\bar{c}$ spectra, rather than experimental states, to determine breakup thresholds for our tetraquark states.

Table 3 presents the lowest "free particle" spectra for $c\bar{c}$ in the first row of masses, the lowest free particle spectra for $c^2\bar{c}^2$ in the second row, and the kinetic energy penalty computed via Eq. (3) in the third row.

The parameters of the Case 1 Hamiltonian are: a=59.4 MeV-fm, m=1490 MeV, $\beta_0 = -1.8\ \mathrm{MeV-fm}^3$, $\beta_1 = 0.3\ \mathrm{MeV-fm}^3$, $\omega = \eta = 0.24\ \mathrm{MeV-fm}^3$, and we set $\sigma = 487.5\ \mathrm{MeV-fm}^{-1}$. The quark mass is 1490 MeV and $\hbar\Omega$ =1280 MeV.

**Table 1: Even orbital parity charmonium masses (MeV) for Case 1. The column N=10 presents the fit to the quoted experimental masses. Other columns present masses obtained in truncated spaces.**

| ID | J | $M_{\exp}$ | N=0 | N=2 | N=4 | N=6 | N=8 | N=10 |
|---|---|---|---|---|---|---|---|---|
| $\eta_c(1S)$ | 0 | 2979 | 3537 | 3369 | 3203 | 3151 | 3092 | 3069 |
| $J/\psi$ | 1 | 3097 | 4302 | 3466 | 3343 | 3198 | 3155 | 3097 |
| $\eta_c(2S)$ | 0 | 3654 | _____ | 4854 | 4417 | 3995 | 3855 | 3676 |
| $\psi(2S)$ | 1 | 3686 | _____ | 5164 | 4609 | 4327 | 4005 | 3878 |

**Table 2: Odd parity charmonium masses (MeV) for Case 1. The column N=9 presents the fit to the quoted experimental masses. Other columns present masses obtained in truncated spaces.**

| ID | J | $M_{\exp}$ | N=1 | N=3 | N=5 | N=7 | N=9 |
|---|---|---|---|---|---|---|---|
| $\chi_{c0}(1P)$ | 0 | 3415 | 4311 | 4014 | 3719 | 3542 | 3423 |
| $\chi_{c1}(1P)$ | 1 | 3511 | 4455 | 4062 | 3808 | 3638 | 3511 |
| $h_c(1P)$ | 1 | 3526 | 4483 | 4071 | 3825 | 3656 | 3528 |
| $\chi_{c2}(1P)$ | 2 | 3556 | 4498 | 4076 | 3834 | 3666 | 3537 |

**Table 3: 2 and 4-body "free particle" spectra with computed correction for Case 1**

| N=0 | N=1 | N=2 | N=3 | N=4 | N=5 | N=6 | N=7 | N=8 | N=9 | N=10 |
|---|---|---|---|---|---|---|---|---|---|---|
| Lowest 2-body free particle spectra (MeV) | | | | | | | | | | |
| 3940 | 4580 | 3568 | 4023 | 3406 | 3761 | 3315 | 3606 | 3256 | 3503 | 3215 |
| Lowest 4-body free particle spectra (MeV) | | | | | | | | | | |
| 8840 | 9480 | 7979 | 8488 | 7538 | 7966 | 7261 | 7632 | 7070 | 7397 | 6929 |
| $\Delta M_{cc\bar{c}\bar{c}}^{free,N,1280}$ (MeV) | | | | | | | | | | |
| 960 | 960 | 471 | 525 | 402 | 375 | 287 | 303 | 258 | 230 | 208 |

Table 4 (5) shows the uncorrected even (odd) parity spectrum for $c^2\bar{c}^2$ in the first block of four rows. The second block shows the spectra, corrected via Eq. (4), obtained by subtracting the kinetic energy penalty from Table 3 from each state for a given N. The corrected masses should not be interpreted as predictions for the masses of the indicated state.

Care is taken in the net binding calculation to conserve parity and total J. In addition, we allow for the tetraquark system to organize into the two energetically most favored $c\bar{c}$ subsystems subject to the limitation of the total tetraquark oscillator quanta available. As an example of how this carried out, let us examine the J=0 state at N=10 in Table 4. We find the most stringent threshold. We easily see that breakup into two odd orbital parity $c\bar{c}$ states is unfavorable. Then we take the minimum mass of two $\eta_c(1S)$ particles that, taken together, have the same total oscillator quanta as the N=10 tetraquark state. We add the N=0 and N=10 $\eta_c(1S)$ masses from Table 1, the N=2 and N=8 masses, and the N=4 and N=6 masses, and simply choose the minimum of these three combinations, the N=4 and N=6 masses in this case. We note this minimum in the last column of Table 4 and similar minima in later tables. In this particular case, the J=0 tetraquark state, the entry "$\eta_c(1S)$, N=4 & 6" indicates that the decay threshold is most stringent with two $\eta_c(1S)$ particles, one evaluated in N=4 and the other in the N=6 basis spaces. This minimum is then subtracted from the corrected J=0 state to estimate the net binding (in parentheses) in Table 4. All thresholds indicated are 1S or 1P states in Tables 4 and 5, as appropriate.

It is apparent that the three lowest states of even parity in Table 4 and all the listed negative parity states in Table 5 are bound, although the binding generally decreases in magnitude for large N. The states at N=9, and the J=2 state at N=10 would barely be bound without the kinetic energy correction of Eq. (4).

Our main focus is on tetraquark binding relative to its most stringent theoretical breakup threshold. Hence, in our own analysis, we employ $c\bar{c}$ results that arise in model spaces with N< $N_{\max}$. However, there is considerable sensitivity of these $c\bar{c}$ masses to the model space dimension. Hence, the corrected tetraquark masses at $N_{\max}$ = 9 (10) in the odd (even) parity should be used for predicting thresholds rather than taken as predictions for the masses of the all-charm tetraquark bound states.

The main import of our results lies in the appreciable binding energies found in the low-lying tetraquark states.

Naturally, we would prefer to carry out this procedure in even larger model spaces (i.e. larger $N_{nax}$) with suitably improved Hamiltonians since the underlying logic for the minimum threshold is designed for asymptotically separated and converged $c\bar{c}$ subsystems. In an expanded treatment with Hamiltonians that produce better descriptions of the experimental $c\bar{c}$ states, we would hope to obtain smooth behavior of both the tetraquark and $c\bar{c}$ spectra with increasing $N_{nax}$. This would yield greater confidence in the predictions, particularly for the tetraquark masses.

**Table 4: Case 1 even parity tetraquarks mass spectra. Raw and corrected spectra, with net binding in parentheses, followed by computed thresholds for N=10.**

| $M_{cc\bar{c}\bar{c}}^{full,N,1280}$ | | | | | | |
|---|---|---|---|---|---|---|
| J | N=0 | N=2 | N=4 | N=6 | N=8 | N=10 |
| 0 | 7422 | 7139 | 6760 | 6615 | 6448 | 6367 |
| 1 | 7896 | 7266 | 6887 | 6682 | 6515 | 6411 |
| 2 | 8252 | 7450 | 6995 | 6762 | 6576 | 6459 |
| 0 | 9615 | 8022 | 7428 | 7095 | 6869 | 6719 |
| $M_{cc\bar{c}\bar{c}}^{corrected,N,1280}$ | | | | | | |
| 0 | 6462<br>(-612) | 6668<br>(-238) | 6358<br>(-380) | 6328<br>(-250) | 6190<br>(-216) | **6159**<br>**(-195)**<br>$\eta_c$<br>N=4 & 6 |
| 1 | 6936<br>(-903) | 6795<br>(-208) | 6485<br>(-350) | 6395<br>(-274) | 6257<br>(-289) | **6203**<br>**(-198)**<br>$\eta_c$, N=4<br>$J/\psi$ ,N=6 |
| 2 | 7292 | 6979 | 6593 | 6475 | 6318 | **6251** |
| | (-1312) | (-789) | (-339) | (-334) | (-346) | **(-290)**<br>$J/\psi$<br>N=4 & 6 |
| 0 | 8655<br>(1581) | 7551<br>(645) | 7026<br>(288) | 6808<br>(230) | 6611<br>(205) | **6511**<br>**(157)**<br>$\eta_c$<br>N=4 & 6 |

**Table 5: Case 1 odd parity tetraquark mass spectra. Raw and corrected spectra, with net binding in parentheses, followed by computed thresholds for N=9.**

| $M_{cc\bar{c}\bar{c}}^{full,N,1280}$ | | | | | |
|---|---|---|---|---|---|
| J | N=1 | N=3 | N=5 | N=7 | N=9 |
| 0 | 8247 | 7752 | 7320 | 7073 | 6876 |
| 1 | 8286 | 7773 | 7343 | 7103 | 6906 |
| 1 | 8329 | 7791 | 7360 | 7113 | 6917 |
| 2 | 8355 | 7800 | 7369 | 7123 | 6926 |
| $M_{cc\bar{c}\bar{c}}^{corrected,N,1280}$ | | | | | |
| 0 | 7287<br>(-561) | 7227<br>(-324) | 6945<br>(-311) | 6770<br>(-309) | **6646**<br>**(-265)**<br>$\eta_c$, N=2 |
| | | | | | $\chi_{co}$, N=7 |
| 1 | 7326<br>(-666) | 7248<br>(-351) | 6968<br>(-377) | 6800<br>(-375) | **6676**<br>**(-331)**<br>$\eta_c$, N=2<br>$\chi_{c1}$, N=7 |
| 1 | 7369<br>(-623) | 7266<br>(-333) | 6985<br>(-360) | 6810<br>(-365) | **6687**<br>**(-320)**<br>$\eta_c$, N=2<br>$\chi_{c1}$, N=7 |
| 2 | 7395<br>(-640) | 7275<br>(-338) | 6994<br>(-377) | 6820<br>(-383) | **6696**<br>**(-339)**<br>$\eta_c$, N=2<br>$\chi_{c2}$, N=7 |

**Case 2 tetraquark results**

Given the number of adjustable parameters in the Hamiltonian, multiple RMS mass difference minima could be obtained from different fits. In order to explore the sensitivity of our results to allowable variations in our Hamiltonian and basis space parameters, we carried out a second fit to the lowest 8 states of the $c\bar{c}$ spectra. The Case 2 parameters for the Hamiltonian are: $\beta_0 = -1.2\ \mathrm{MeV-fm}^3$, $\beta_1 = 0.45\ \mathrm{MeV-fm}^3$, $\omega = 0.225\ \text{MeV} - \text{fm}^3$, $\eta = 0.30\ \text{MeV} - \text{fm}^3$, and we have $\hbar\Omega$ =1200 MeV. Note that the Coulomb-like interaction coupling and the confining strength are unchanged.

We present the results of a fit to the low-lying even and odd orbital parity spectrum of charmonium in Tables 6 and 7. The RMS mass difference for even orbital parity at $N_{\max}$=10 is 87.0 (62.7) MeV with (without) the $\psi(2S)$. The RMS mass difference for the odd orbital parity at $N_{\max}$=9 is 11 MeV. The RMS mass difference for all 8 states is 62.0 MeV. Case 2 provides a better overall fit to the charmonium spectrum, but if one omits the $\psi(2S)$, Case 1 yields a slightly better fit. The interpretation of Tables 6 through 10 follows the same path as Tables 1 through 5 for Case 1.

The tetraquark binding is significantly less than Case 1. Looking at the important relative S-state interaction components, we observe that Case 2 has less attractive delta function strengths and a less repulsive tensor term. The decreased tetraquark binding may well be correlated with the S-state properties of our effective interaction. A more extensive exploration of this sensitivity is needed to establish this correlation.

The "kinetic energy deficit" in Table 8 for large N is very similar to Case 1, which is expected since the deficit depends only on the basis space parameters and the quark mass. The uncorrected masses in Tables 9 and 10 for the tetraquark are about 100 MeV higher than Case 1 results, which leads to the reduced binding since the decay thresholds are virtually unchanged. We note that, unlike Case 1, none of the states would be bound at $N_{\max}$=9 or 10 if the kinetic energy correction was omitted.

**Table 6: Even parity charmonium masses (MeV) for Case 2. The column N=10 presents the fit to the quoted experimental masses. Other columns present masses obtained in truncated spaces.**

| ID | J | $M_{\exp}$ | N=0 | N=2 | N=4 | N=6 | N=8 | N=10 |
|---|---|---|---|---|---|---|---|---|
| $\eta_c(1S)$ | 0 | 2979 | 4024 | 3352 | 3187 | 3152 | 3089 | 3074 |
| $J/\psi$ | 1 | 3097 | 4024 | 3442 | 3320 | 3198 | 3155 | 3105 |
| $\eta_c(2S)$ | 0 | 3654 | _____ | 4637 | 4301 | 3884 | 3783 | 3602 |
| $\psi(2S)$ | 1 | 3686 | _____ | 5053 | 4545 | 4268 | 3943 | 3822 |

**Table 7: Odd parity charmonium masses (MeV) for Case 2. The column N=9 presents the fit to the quoted experimental masses. Other columns present masses obtained in truncated spaces.**

| ID | J | $M_{\exp}$ | N=1 | N=3 | N=5 | N=7 | N=9 |
|---|---|---|---|---|---|---|---|
| $\chi_{c0}(1P)$ | 0 | 3415 | 4250 | 3982 | 3704 | 3536 | 3422 |
| $\chi_{c1}(1P)$ | 1 | 3511 | 4384 | 4028 | 3788 | 3625 | 3503 |
| $h_c(1P)$ | 1 | 3526 | 4417 | 4039 | 3808 | 3647 | 3523 |
| $\chi_{c2}(1P)$ | 2 | 3556 | 4440 | 4046 | 3822 | 3662 | 3537 |

**Table 8: 2 and 4-body free particle spectra with computed correction.**

| N=0 | N=1 | N=2 | N=3 | N=4 | N=5 | N=6 | N=7 | N=8 | N=9 | N=10 |
|---|---|---|---|---|---|---|---|---|---|---|
| Lowest 2-body free particle spectra (MeV) | | | | | | | | | | |
| 3900 | 4500 | 3551 | 3977 | 3400 | 3732 | 3314 | 3587 | 3259 | 3491 | 3220 |
| Lowest 4-body free particle spectra (MeV) | | | | | | | | | | |
| 8700 | 9300 | 7893 | 8370 | 7479 | 7880 | 7220 | 7567 | 7041 | 7347 | 6908 |
| $\Delta M_{cc\bar{c}\bar{c}}^{free,N,1200}$ (MeV) | | | | | | | | | | |
| 900 | 900 | 442 | 493 | 377 | 352 | 269 | 284 | 241 | 215 | 194 |

**Table 9: Case 2 even parity tetraquarks mass spectra. Raw and corrected spectra, with net binding in parentheses, followed by computed thresholds for N=10.**

| $M_{cc\bar{c}\bar{c}}^{full,N,1200}$ | | | | | | |
|---|---|---|---|---|---|---|
| J | N=0 | N=2 | N=4 | N=6 | N=8 | N=10 |
| 0 | 7519 | 7196 | 6854 | 6700 | 6555 | 6477 |
| 1 | 7870 | 7320 | 6974 | 6774 | 6624 | 6528 |
| 2 | 8146 | 7458 | 7068 | 6845 | 6680 | 6573 |
| 0 | 8941 | 7922 | 7315 | 7026 | 6829 | 6695 |
| | | | $M_{cc\bar{c}\bar{c}}^{corrected,N,1200}$ | | | |
| 0 | 6619<br>(-1429) | 6754<br>(-622) | 6477<br>(-227) | 6431<br>(-108) | 6314<br>(-60) | **6283**<br>**(-56)**<br>$\eta_c$,<br>N=4 & 6 |
| 1 | 6970<br>(-1078) | 6878<br>(-498) | 6597<br>(-197) | 6505<br>(-124) | 6383<br>(-124) | **6334**<br>**(-51)**<br>$\eta_c$, N=4<br>$J/\psi$ ,N=6 |
| 2 | 7246<br>(-802) | 7016<br>(-430) | 6691<br>(-193) | 6576<br>(-186) | 6439<br>(-201) | **6379**<br>**(-147)**<br>$J/\psi$<br>N=4 & 6 |
| 0 | 8041<br>(-7) | 7350<br>(-26) | 6938<br>(234) | 6757<br>(218) | 6588<br>(214) | **6501**<br>**(162)**<br>$\eta_c$,<br>N=4 & 6 |

**Table 10: Case 2 odd parity tetraquark mass spectra. Raw and corrected spectra, with net binding in parentheses, followed by computed thresholds for N=9.**

| $M_{cc\bar{c}\bar{c}}^{full,N,1200}$ | | | | | |
|---|---|---|---|---|---|
| J | N=1 | N=3 | N=5 | N=7 | N=9 |
| 0 | 8310 | 7806 | 7400 | 7154 | 6969 |
| 1 | 8363 | 7830 | 7429 | 7190 | 7004 |
| 1 | 8388 | 7843 | 7440 | 7197 | 7013 |
| 2 | 8420 | 7856 | 7454 | 7214 | 7033 |
| $M_{cc\bar{c}\bar{c}}^{corrected,N,1200}$ | | | | | |
| 0 | 7410<br>(-864) | 7313<br>(-289) | 7048<br>(-286) | 6870<br>(-186) | **6754**<br>**(-134)**<br>$\eta_c$, N=2<br>$\chi_{co}$, N=7 |
| 1 | 7463<br>(-811) | 7337<br>(-355) | 7077<br>(-303) | 6906<br>(-234) | **6789**<br>**(-186)**<br>$\eta_c$, N=4<br>$\chi_{c1}$, N=5 |
| 1 | 7488<br>(-786) | 7350<br>(-342) | 7088<br>(-292) | 6913<br>(-227) | **6798**<br>**(-177)**<br>$\eta_c$, N=4<br>$\chi_{c1}$, N=5 |
| 2 | 7520<br>(-888) | 7363<br>(-429) | 7102<br>(-296) | 6930<br>(-244) | **6818**<br>**(-191)** |

| | | | | | | |
|---|---|---|---|---|---|---|
| | | | | | $\eta_c$, N=4<br>$\chi_{c2}$, N=5 | |

## Summary and Conclusions

We have used a parameterized Hamiltonian to compute the spectrum of all-charm tetraquark states. After fitting the lowest $c\bar{c}$ states with two different parameter sets and performing a net binding analysis, we obtain bound tetraquark states with both sets of parameters. For example, the lowest tetraquark state with J=0 in the positive parity spectrum has a mass below the threshold of two $\eta_c(1S)$ masses computed in our framework.

A simple extrapolation of the binding energies might lead one to suspect some states would become unbound with increasing basis space size, $N_{\max}$. However, the trends in the binding energies are not smooth functions of increasing N and make extrapolation a risky proposition. Furthermore, simple extrapolation is not warranted since the Hamiltonian parameters were fit within the $N_{\max}$=9 (10) basis spaces to the negative (positive) orbital parity experimental meson masses.

Experimental investigations could identify the states we predict at $e^+e^-$ facilities. Indeed, Iwasaki[20] originally proposed measuring the recoil-mass spectrum in coincidence with a $J/\psi$ around CM energy of 6 GeV to look for all-charm tetraquark resonances. Production of these tetraquark states, however, may be less favorable at modest CM energies. In fact, there has been theoretical speculation[21] that certain perturbative diagrams for $J/\psi c\bar{c}$ production in $e^+e^-$ annihilation may be enhanced at CM energies greater than 10 GeV, where Belle and BaBar have already reported double charmonium production. This leads us to speculate that searching for tetraquark resonances at higher CM energies may be more fruitful than searches around 6 GeV.

In a further extension of this work, we intend to employ a Hamiltonian with broader phenomenological success[22] that includes relativistic kinematics effects as well as a treatment of negative frequency states.

## Acknowledgements

This work was supported in part by U. S. DOE Grant No. DE-FG-02-87ER-40371, Division of High Energy and Nuclear Physics. The authors also acknowledge valuable discussions with J. R. Spence and Alaa Abd El-Hady, who particularly contributed to the early stages of this work. We would also like to thank Chris Quigg for pointing out the existence of Iwasaki's work.

[1] Y. Iwasaki, Prog. Theor. Phys. **54**, 492 (1975).

[2] R. J. Jaffe, Phys. Rev. D **15**, 267 (1977).

[3] J. Schaffner-Bielich and A. P. Vischer, Phys. Rev. D **57**, 4142 (1998).

[4] J. Weinstein and N. Isgur, Phys. Rev. Lett. **48**, 569 (1982).

[5] Stephen Godfrey and Nathan Isgur, Phys. Rev. D **32**, 189 (1985).

[6] T. E. O. Ericson and G. Karl, Phys. Lett. B **309**, 426 (1993).

[7] Fl. Stancu, Talk presented at the International Workshop on Hadron Physics "Effective Theories of Low Energy QCD", Coimbra, Portugal, Sep. 10-15, 1999.

[8] L. Ya. Glozman and D. O. Riska, Phys. Rep. **268**, 263 (1996); L. Ya. Glozman, Z Papp, and W. Plessas, Phys. Lett. **B381**, 311 (1996).

[9] Gunnar S. Bali, Phys. Rept. **343**, 1 (2001).

[10] B. Aubert et al., Phys. Rev. Lett. **90**, 242001 (2003).

[11] T. Barnes, F. E. Close, and H. J. Lipkin, Phys. Rev. D **68**, 054006 (2003).

[12] K. Abe, et. al, Phys. Rev. Lett. **89**, 142001 (2002).

[13] B. Aubert, et. al., Phys. Rev. Lett **87**, 162002 (2001).

[14] R. S. Willey, Phys. Rev. D **18**, 270 (1978).

[15] O. W. Greenberg and Harry J. Lipkin, Nucl. Phys. A **370**, 349 (1981).

[16] J. P. Vary, The Many-Fermion Dynamics Shell-Model Code, Iowa State University (1992) (unpublished)

[17] P. Navratil, J. P. Vary, and B. R. Barrett, Phys. Rev. C **62**, 054311 (2000).

[18] K. Hagiwara, et. al., Phys. Rev. D **66**, 010001 (2002).

[19] Horace Crater and Peter Van Alstine, hep-ph/0208186

[20] Yoichi Iwasaki, Phys. Rev. D **16**, 220 (1977).

[21] B. L. Joffe and D. E. Kharzeev, arXiv: hep-ph/0306062

[22] J. R. Spence and J. P. Vary, Phys. Rev. C **59**, 1762 (1999).